%% file: main.tex
\documentclass{article}

\usepackage[preprint]{neurips_2026}

\usepackage[utf8]{inputenc}
\usepackage[T1]{fontenc}
\usepackage{hyperref}
\usepackage{url}
\usepackage{booktabs}
\usepackage{amsfonts}
\usepackage{amsmath}
\usepackage{amssymb}
\usepackage{amsthm}
\usepackage{nicefrac}
\usepackage{microtype}
\usepackage{xcolor}
\usepackage{graphicx}
\usepackage{subcaption}
\usepackage{algorithm}
\usepackage{algorithmic}
\usepackage{multirow}

\newcommand{\R}{\mathbb{R}}

\title{Not All Memories Age the Same: Autodiscovery of Adaptive Decay in Knowledge Graphs}

\author{
  Mandar Karhade \\
  \texttt{mandar.karhade@citingale.com}
}

\begin{document}

\maketitle

\input{sections/abstract}
\input{sections/introduction}
\input{sections/related_work}
\input{sections/framework}
\input{sections/inference}
\input{sections/experiments}
\input{sections/analysis}
\input{sections/discussion}
\input{sections/conclusion}

\bibliographystyle{plain}
\bibliography{bibliography}

\newpage
\appendix
\part*{Appendix}
\input{sections/appendix}

\end{document}

%% file: sections/abstract.tex
\begin{abstract}

Knowledge graphs used for retrieval treat all facts as equally current. Existing temporal approaches apply uniform decay, using a single forgetting curve regardless of knowledge type. We show this is fundamentally misspecified: different knowledge types exhibit different temporal dynamics, and the core retrieval problem is not latency or throughput but \emph{identifying what is important at query time}.

We propose a hierarchical framework that replaces uniform decay with a continuous decay surface parameterized by two orthogonal signals: \textbf{velocity} (how frequently a concept is observed) and \textbf{volatility} (how much the value changes between observations, measured via embedding distance). The decay surface is decomposed into three learnable levels: \emph{domain-level} parameters capture universal patterns, \emph{context-level} parameters capture setting-dependent variation, and \emph{entity-level} adaptation personalizes decay to specific subjects. All parameters emerge from data through survival analysis on observed value lifetimes, requiring no predefined taxonomies or domain expertise.

We formulate edge lifetime as a survival problem where the event is \emph{value supersession} (a meaningfully different value replacing the current one), distinct from mere re-observation.

Experiments on synthetic temporal knowledge graphs demonstrate recovery of planted hierarchical parameters (HDBSCAN ARI = 1.0). Validation on 107 Wikipedia articles and 1{,}163 patient records from the Synthea clinical EHR simulator shows that velocity-volatility clusters emerge naturally, align with observable persistence patterns, and near-universally exhibit the Lindy effect (Weibull shape $\kappa < 1$): older facts are less likely to be superseded, with a single clinically coherent exception (chemotherapy drugs, which exhibit aging). Uniform decay performs 18$\times$ worse than no temporal weighting. Heterogeneous decay recovers from this, with each hierarchy level contributing measurable improvement.

\end{abstract}

%% file: sections/introduction.tex
\section{Introduction}
\label{sec:introduction}

The central challenge of retrieval from knowledge graphs is determining what is relevant at query time. This requires more than semantic matching: the system must also assess whether a retrieved edge is still current, and that assessment depends on the nature of the knowledge itself. A genetic test result from five years ago may be perfectly current while a blood pressure reading from yesterday may already be stale. No single decay rate can accommodate both.

This problem is compounded by three specific challenges that existing systems fail to address:

\textbf{1. Temporal currency depends on knowledge type.} A uniform decay function, the standard approach in temporal retrieval systems \cite{zhong2024memorybank, park2023generative}, applies the same forgetting curve to all edges. Consider a clinical knowledge graph queried in March 2025 for a melanoma patient. The graph contains a BRAF V600E mutation documented in 2021, a treatment regimen from 2023 since replaced, a CT scan finding from 2023 since contradicted by disease progression, and the current treatment started in 2024. Uniform decay simultaneously suppresses the mutation (still true, penalized for age) and retains the scan finding (no longer true, not penalized enough). The failure is not in the decay rate but in the assumption that all knowledge ages uniformly.

\textbf{2. Infrequent mention does not mean low relevance.} In clinical records, some of the most critical findings appear rarely. A genetic mutation may be documented once in a patient's entire history, yet it is the single most relevant fact for treatment decisions at any future query time. A blood pressure reading appears thousands of times, yet each individual reading is relevant for only hours. Systems that use mention frequency as a proxy for importance systematically bury rare but permanent knowledge under frequent but transient observations.

\textbf{3. Concepts and values have different temporal properties.} The concept ``blood pressure'' is always relevant to a patient's status; it never expires. But a specific blood pressure \emph{value} (120/80 measured last Tuesday) has a short shelf life. Similarly, the concept ``treatment regimen'' is always important to retrieve, but the specific value (pembrolizumab started in January 2023) becomes stale when the treatment changes. A useful information model must separate the enduring relevance of a concept from the temporal currency of its value.

These three challenges motivated the development of a framework that accommodates temporality through a learned hierarchy, assigns different shelf lives to different knowledge types, and distinguishes concepts from their values.

\subsection{Existing Approaches and Their Limitations}

Existing temporal approaches fall into two categories. Temporal knowledge graph models (RE-GCN \cite{li2021regcn}, TRCL, TiRGN) learn temporal embeddings for link prediction, that is, predicting future facts, but do not address retrieval weighting of existing facts. Memory decay systems (MemoryBank \cite{zhong2024memorybank}, Generative Agents \cite{park2023generative}, FOREVER \cite{feng2026forever}) apply Ebbinghaus-inspired uniform decay, sometimes modulated by a static importance score set at storage time. Neither decomposes temporal dynamics by knowledge type, and neither learns decay parameters from observed edge behavior.

\subsection{Our Approach}

We address these challenges through two orthogonal signals and a hierarchical model:

\textbf{Velocity and volatility.} We decompose temporal behavior into two independent measurements. \emph{Velocity} captures how frequently a concept is observed: blood pressure is measured daily in intensive care, quarterly in outpatient settings, while a genetic test may be performed once in a patient's lifetime. \emph{Volatility} captures how much the value changes between consecutive observations, measured as embedding distance in a shared representation space. Crucially, these are kept orthogonal: volatility is not normalized by observation interval, so a concept can be high-velocity but low-volatility (frequently measured but stable) or low-velocity but high-volatility (rarely measured but changing each time).

This decomposition directly addresses challenge 2: velocity reflects observation patterns, not importance. A genetic mutation with low velocity and zero volatility receives a long shelf life, while a vital sign with high velocity and high volatility receives a short one. Frequency of mention is decoupled from temporal relevance.

\textbf{Value supersession.} We formulate edge lifetime as a survival problem where the event of interest is \emph{value supersession}: a new observation with a meaningfully different value, not merely a new observation. A blood pressure re-reading of 120/80 following a prior 120/80 is a \emph{reinforcement} (confirming the value is still current), not a supersession. A reading of 145/95 following 120/80 is a supersession (the old value is now stale). This directly addresses challenge 3: the concept persists while the value's shelf life is determined by when it is superseded.

\textbf{Continuous decay surface.} Rather than discretizing knowledge into categories (``permanent'' versus ``transient''), we model the shelf life $\tau$ as a continuous function of velocity and volatility: $\tau(v, \sigma) = \exp(\theta_0 + \theta_1 v + \theta_2 \sigma + \theta_3 v\sigma)$. Permanence is not a category; it is a position where velocity and volatility are both near zero and the function naturally evaluates to a very long shelf life.

\textbf{Hierarchical decomposition.} The decay surface parameters decompose into multiple learnable levels: domain-level (universal patterns across a specialty or knowledge area), context-level (setting-dependent modulation), and entity-level (adaptation to a specific patient, article, or case). The hierarchy is Bayesian: new entities borrow strength from their context, new contexts from the domain prior. As data accumulates, lower-level parameters dominate. This directly addresses challenge 1: the decay parameters are not assumed but learned from how knowledge actually behaves over time, at multiple levels of granularity.

\subsection{Contributions}

\begin{enumerate}
    \item An information model that separates concept relevance from value currency, using velocity (observation frequency) and volatility (value change magnitude) as orthogonal signals, with a continuous decay surface replacing uniform decay or discrete taxonomies.
    \item A hierarchical Bayesian survival model where domain, context, and entity-level shelf life parameters are learned from observed value lifetimes without domain expertise.
    \item An embedding-based volatility measure that handles numeric, categorical, and free-text values uniformly, enabling the framework to operate across domains without type-specific preprocessing.
    \item Empirical validation on synthetic data, Wikipedia revision history, and Synthea clinical EHR records showing that uniform decay performs 18$\times$ worse than no temporal weighting, while heterogeneous decay recovers and improves precision. Temporal clusters emerge from data and align with domain intuition. The Lindy effect ($\kappa < 1$) holds near-universally (all Wikipedia clusters, 9 of 10 Synthea clusters), with the single clinically coherent exception (chemotherapy regimens) discovered automatically by the framework.
\end{enumerate}

%% file: sections/related_work.tex
\section{Related Work}
\label{sec:related_work}

\subsection{Temporal Knowledge Graphs}

Temporal knowledge graph (TKG) models capture the dynamic nature of facts that evolve over time. RE-GCN \cite{li2021regcn} learns evolutional representations via graph convolutional networks on temporal snapshots. A comprehensive survey by \cite{cai2023survey} covers representation learning approaches for TKGs.

These methods focus on \emph{link prediction} (forecasting future facts from historical patterns). Our work addresses a complementary problem: \emph{retrieval weighting} of existing facts at query time. TKG models predict what will happen; we score what is still true.

\subsection{Memory Decay in LLM Systems}

MemoryBank \cite{zhong2024memorybank} applies the Ebbinghaus forgetting curve to conversational memory, modeling memory strength as a discrete value that increases with recall and decreases over time. Generative Agents \cite{park2023generative} score memories by recency $\times$ importance $\times$ relevance, where importance is static and set at creation time. FOREVER \cite{feng2026forever} applies forgetting curves to continual learning replay scheduling, addressing the challenge of defining meaningful ``time'' for LLMs.

All of these systems apply uniform or static-importance decay. None decompose temporal dynamics into setting-aware, context-dependent, or entity-adapted components. None learn decay parameters from observed edge behavior; the decay function is assumed, not fitted.

\subsection{Temporal Information Retrieval}

TCR-QF \cite{huang2025tcrqf} restores missing temporal context in knowledge graphs during reasoning through query-driven feedback. Existing temporal approaches in retrieval-augmented generation apply heuristic recency priors, such as half-life decay functions fused with semantic similarity. These approaches observe that retrieval pipelines without temporal components risk elevating stale information, but apply uniform decay rates regardless of knowledge type.

Our framework provides a principled, learned alternative where decay parameters are heterogeneous (different for different knowledge types) and emerge from survival analysis on observed value lifetimes.

\subsection{Survival Analysis}

Cox proportional hazards models \cite{cox1972regression} are the standard tool for modeling time-to-event data as a function of covariates. Bayesian hierarchical survival models, particularly frailty models \cite{ibrahim2001bayesian}, introduce random effects for entity-level variation, directly analogous to our entity-level parameters.

We bring survival analysis methodology to knowledge graph temporal management. The statistical framework is well-established \cite{kleinbaum2012survival}; the \emph{application} to KG retrieval weighting, with the concept of value supersession as the survival event and velocity-volatility as covariates, is novel.

\subsection{Hierarchical Bayesian Models}

Hierarchical Bayesian models \cite{gelman2013bayesian, gelman2006data} are standard in biostatistics for borrowing strength across groups. Our contribution is not the Bayesian machinery but its application to temporal KG dynamics, with the specific hierarchy (domain $\to$ context $\to$ entity) and the connection between velocity-volatility covariates and survival parameters.

%% file: sections/framework.tex
\section{Framework}
\label{sec:framework}

We present the framework in three parts. Section~\ref{sec:definitions} defines the signals and events that the model operates on. Section~\ref{sec:structure} describes how knowledge is organized into clusters and hierarchical levels. Section~\ref{sec:estimation} presents the statistical machinery for estimating shelf life parameters and applying them at retrieval time.

\subsection{Definitions}
\label{sec:definitions}

\subsubsection{Preliminaries and Notation}

A temporal knowledge graph $\mathcal{G} = \{(s, p, o, t)\}$ consists of edges $e = (s, p, o, t)$ where $s$ is the subject, $p$ is the predicate, $o$ is the object, and $t$ is the extraction timestamp. We use $e.s$, $e.p$, $e.o$, $e.t$ to denote the components of an edge. We introduce four terms used throughout.

\textbf{Predicate} ($p$) is the relationship type, independent of any specific subject. Examples include ``blood\_pressure'', ``treatment'', and ``BRAF\_status''. All edges sharing the same predicate describe the same kind of knowledge.

\textbf{Concept} ($c = (s, p)$) is a subject-predicate pair: ``Patient A's blood pressure'' or ``Patient A's treatment''. A concept is a specific measurement or attribute of a specific entity.

\textbf{Value} ($o$) is the object of an edge, representing the specific instantiation of a concept at time $t$. Examples: ``120/80'', ``pembrolizumab'', ``V600E''.

\textbf{Shelf life} ($\tau$) is the characteristic time until a value becomes stale, that is, until it is likely to be superseded by a meaningfully different value. A short shelf life (hours) means the value expires quickly; a long shelf life (years) means it remains current indefinitely.

Predicates and concepts play different roles in the framework. Velocity and volatility are measured per \emph{concept} (Patient A's blood pressure has a specific observation frequency and value stability). Clustering operates on \emph{predicates} (blood\_pressure as a type behaves similarly across all patients). The hierarchy bridges the two: Level~1 groups predicates by temporal behavior, while Level~3 personalizes to individual concepts.

\subsubsection{Velocity and Volatility}
\label{sec:velocity_volatility}

We define two orthogonal signals that characterize the temporal behavior of knowledge.

\textbf{Velocity} of a concept $c = (s, p)$ at time $t$ is the rate of new observations per unit time:

\begin{equation}
    \text{vel}(c, t) = \frac{|\{e \in \mathcal{G} : e.s = s, e.p = p, e.t \in [t - \Delta, t]\}|}{\Delta}
\end{equation}

where $\Delta$ is a lookback window. Velocity reflects the observation cadence of a concept, driven by workflow, monitoring protocols, or reporting frequency.

\textbf{Volatility} of a concept $c = (s, p)$ is the mean embedding distance between consecutive observed values:

\begin{equation}
    \text{vol}(c) = \frac{1}{m-1} \sum_{i=1}^{m-1} d\big(\phi(o_i), \phi(o_{i+1})\big)
\end{equation}

where $o_1, \ldots, o_m$ are the $m$ values observed for concept $c$ at successive timestamps, $\phi: \mathcal{V} \to \R^d$ is an embedding function mapping values of any type (numeric, categorical, free text) to a shared vector space, and $d(\cdot, \cdot)$ is a distance metric in the embedding space.

Crucially, volatility is \emph{not} normalized by observation interval. Doing so would conflate volatility with velocity, violating their orthogonality. Volatility measures how much the value changes \emph{when observed}; velocity measures how often observations occur. These are independent phenomena.

\subsubsection{Value Supersession}
\label{sec:supersession}

We model edge lifetime as a survival problem. The event of interest is \emph{value supersession}: the appearance of a new observation with a meaningfully different value for the same concept.

An edge $e = (s, p, o, t)$ is superseded by edge $e' = (s, p, o', t')$ if:

\begin{equation}
    t' > t \quad \text{and} \quad d\big(\phi(o), \phi(o')\big) > \epsilon_p
\end{equation}

where $\epsilon_p$ is a threshold specific to predicate $p$, defining what constitutes a ``meaningful'' change for this kind of knowledge. The lifetime of edge $e$ is then $T_e = t' - t$.

This produces three types of observations. \textbf{Superseded} edges have a subsequent observation with a meaningfully different value; the lifetime $T_e$ is observed exactly. \textbf{Active} edges have not been superseded; the lifetime is right-censored at $T_e^{+} = t_{\text{now}} - t$. \textbf{Reinforced} edges have a subsequent observation with the same value (embedding distance below $\epsilon_p$); this confirms the edge is still alive and resets the observation clock.

\subsection{Structure: Clusters and Hierarchy}
\label{sec:structure}

\subsubsection{Hierarchical Clustering}
\label{sec:clustering_hierarchy}

We organize knowledge into a hierarchy of three levels. Level~1 is discovered through unsupervised clustering on temporal behavior. Levels~2 and~3 are derived through variance decomposition within the clusters established at Level~1.

\textbf{Level 1: Domain Clustering.} The first structural question is: what kinds of knowledge exist in the graph? We answer this by clustering \emph{predicates} (not individual concepts) based on their observed temporal behavior, aggregated across all subjects. For each predicate with sufficient observation history, we compute a feature vector:

\begin{equation}
    \mathbf{x}_p = \big[\text{vel}(p),\; \text{vol}(p),\; \bar{T}_p,\; \rho_p,\; \phi(p)\big]
\end{equation}

where $\bar{T}_p$ is the mean observed lifetime for edges with predicate $p$, $\rho_p$ is the fraction of re-observations that involved a value change, and $\phi(p)$ is the predicate embedding (for cold-start assignment of new predicates).

Predicates with similar temporal profiles cluster together. ``Blood pressure'' and ``heart rate'' (both high-velocity, high-volatility) land in the same cluster. ``BRAF mutation status'' and ``blood type'' (both low-velocity, zero-volatility) land in another. We use HDBSCAN \cite{campello2013hdbscan} and Dirichlet Process Gaussian Mixture Models (DPGMM) \cite{blei2006variational}, comparing both in our experiments. No domain labels or predefined taxonomies are used; the types emerge from observed behavior. Each discovered cluster represents a temporal knowledge type with a characteristic shelf life profile, parameterized by $\theta_k$.

\textbf{Level 2: Context (via Variance Decomposition).} Within each domain cluster, the shelf life of a specific edge varies depending on the setting in which it was observed. A blood pressure reading has a shelf life of hours in an ICU but months in an outpatient clinic. The knowledge type is identical; the setting modulates how quickly the value becomes stale.

We capture this variation by decomposing the within-cluster variance into context-level groups. Context parameters are drawn from the domain prior:

\begin{equation}
    \theta_{k,c} \sim \mathcal{N}(\theta_k, \sigma^2_{\text{context}} I)
    \label{eq:context}
\end{equation}

Contexts may be predefined (care settings, market segments, jurisdictions) or discovered by grouping entities whose edges exhibit similar shelf life deviations from the domain baseline. The variance $\sigma^2_{\text{context}}$ quantifies how much the context shifts the shelf life relative to the domain average.

\textbf{Level 3: Entity (via Variance Decomposition).} Within each context, individual entities deviate from the context baseline. A septic patient in the ICU has blood pressure values that expire in minutes, while a stable post-surgical patient in the same ICU has values that last hours. Entity parameters are drawn from the context prior:

\begin{equation}
    \theta_{k,c,i} \sim \mathcal{N}(\theta_{k,c}, \sigma^2_{\text{entity}} I)
    \label{eq:entity}
\end{equation}

The variance $\sigma^2_{\text{entity}}$ quantifies how much individual entities deviate from their context peers. This deviation is itself informative: an entity whose values turn over faster than expected may be under different conditions, without explicitly modeling those conditions.

The framework is not fixed at three levels. The Bayesian formulation generalizes to any depth: each additional level introduces parameters drawn from the level above, with the same variance decomposition and regularization mechanism.

\subsubsection{Cold Start}
\label{sec:cold_start}

The hierarchy naturally handles sparse data by borrowing strength from higher levels. A new entity with no history uses its context's parameters. A new context uses the domain cluster prior. A new predicate is assigned to a cluster via a classifier trained on predicate embeddings. As observations accumulate, lower-level parameters dominate through Bayesian updating. The transition from prior-dominated to data-dominated estimates is smooth and automatic.

\subsection{Estimation and Retrieval}
\label{sec:estimation}

\subsubsection{Decay Surface}
\label{sec:decay_surface}

Rather than discretizing velocity and volatility into categories, we model the characteristic shelf life $\tau$ as a continuous function:

\begin{equation}
    \tau(v, \sigma) = \exp\big(\theta_0 + \theta_1 v + \theta_2 \sigma + \theta_3 v\sigma\big)
    \label{eq:tau_surface}
\end{equation}

where $v = \text{vel}(c)$ is the velocity and $\sigma = \text{vol}(c)$ is the volatility of the concept. The parameters $\theta = (\theta_0, \theta_1, \theta_2, \theta_3)$ control the surface shape. $\theta_0$ sets the baseline log-shelf-life. $\theta_1$ and $\theta_2$ capture the effects of velocity and volatility on shelf life (both signs and magnitudes are learned; $\theta_2$ is expected to be negative since higher volatility implies shorter shelf life). $\theta_3$ captures the interaction between the two signals.

The $\theta$ parameters are what the hierarchy decomposes. At Level~1, $\theta_k$ defines the decay surface for cluster $k$. At Level~2, $\theta_{k,c}$ shifts the surface for context $c$. At Level~3, $\theta_{k,c,i}$ further adjusts for entity $i$. Permanence is not a special case: an edge with $v \approx 0$ and $\sigma \approx 0$ naturally obtains a very large $\tau$, learned from the data.

\subsubsection{Context-Calibrated Floor}
\label{sec:tau_floor}

The decay surface can produce arbitrarily small $\tau$ values for high-velocity, high-volatility predicates. However, a shelf life shorter than the observation cadence is meaningless: a value cannot be considered stale before the next observation could have arrived. For predicates that describe discrete events rather than persistent states (e.g., ``patient received epinephrine'' versus ``patient's blood pressure is 120/80''), the decay surface may estimate $\tau \approx 0$, producing degenerate shelf lives.

We address this by imposing a learned floor derived from the context-level observation cadence:

\begin{equation}
    \tau_{\text{eff}}(k, c, i) = \max\big(\tau(v, \sigma \mid \theta_{k,c,i}),\; \tau_{\text{floor}}(k, c)\big)
\end{equation}

where $\tau_{\text{floor}}(k, c)$ is the median inter-observation interval for all predicates in cluster $k$ within context $c$. This floor is not a manually chosen parameter. It is computed from the data and reflects the operational cadence of the context: how frequently this type of knowledge is re-observed in this setting.

The floor is set at the context level (Level~2) rather than the predicate level because predicates within the same clinical panel or treatment regimen share an observation cadence. A lab test ordered as part of a comprehensive metabolic panel should not receive a shorter floor just because 13 other tests were ordered simultaneously. The context-level cadence correctly captures the rate at which new information of this type arrives.

\subsubsection{Survival Distribution}
\label{sec:survival_dist}

Given the hierarchical parameters and the context-calibrated floor, the lifetime of edge $e$ belonging to entity $i$, context $c$, and domain cluster $k$ is:

\begin{equation}
    T_e \sim \text{Weibull}\big(\tau_{\text{eff}}(k, c, i),\; \kappa_k\big)
\end{equation}

where $\kappa_k$ is the shape parameter shared within domain cluster $k$. The Weibull distribution generalizes exponential decay. When $\kappa_k = 1$, the hazard is constant (memoryless, equivalent to exponential). When $\kappa_k > 1$, older values are more likely to change (an aging effect). When $\kappa_k < 1$, older values are less likely to change (a Lindy effect \cite{taleb2012antifragile}). Whether knowledge exhibits Lindy-like behavior or aging is an empirical question we investigate in Section~\ref{sec:experiments}.

We choose the Weibull family for its interpretable shape parameter and its generalization of exponential decay. Empirically, log-normal distributions often provide a better fit to observed edge lifetimes (Section~\ref{sec:ablations}). This is expected: a hierarchical mixture of Weibulls (edges with different entity-level $\tau$ values drawn from the same cluster) produces a marginal distribution that approximates log-normal. The Weibull parameterization remains preferable because $\kappa_k$ directly encodes the aging-versus-Lindy distinction per cluster, which is a primary quantity of interest.

The survival distribution is fitted at \emph{every level} of the hierarchy, not only at Level~1. At each level, the Weibull parameters are estimated from the edges belonging to that group (cluster, context, or entity), with regularization toward the level above. Survival fitting and hierarchical decomposition are not separate steps but a single integrated estimation process.

\subsubsection{Likelihood}
\label{sec:likelihood}

The full likelihood over all edges, given parameters $\Theta = \{\theta_{k,c,i}, \kappa_k\}$, is:

\begin{equation}
    \mathcal{L}(\Theta) = \prod_{e \in \mathcal{S}} f(T_e \mid \Theta_e) \cdot \prod_{e \in \mathcal{A}} S(t_{\text{now}} - t_e \mid \Theta_e) \cdot \prod_{e \in \mathcal{R}} S(t_r^{(e)} - t_e \mid \Theta_e)
    \label{eq:likelihood}
\end{equation}

where $\mathcal{S}$ is the set of superseded edges, $\mathcal{A}$ is the set of active (right-censored) edges, $\mathcal{R}$ is the set of reinforced edges, $f(\cdot)$ is the Weibull density, and $S(\cdot)$ is the Weibull survival function:

\begin{align}
    f(t \mid \tau, \kappa) &= \frac{\kappa}{\tau} \left(\frac{t}{\tau}\right)^{\kappa - 1} \exp\left(-\left(\frac{t}{\tau}\right)^{\kappa}\right) \\
    S(t \mid \tau, \kappa) &= \exp\left(-\left(\frac{t}{\tau}\right)^{\kappa}\right)
\end{align}

\subsubsection{Temporal Retrieval Scoring}
\label{sec:retrieval_scoring}

At query time with timestamp $t_q$, the retrieval score for edge $e$ combines semantic relevance with temporal freshness:

\begin{equation}
    \text{score}(e, q, t_q) = \text{sim}(e, q)^{\alpha} \cdot S(t_q - t_e \mid \tau(v_e, \sigma_e),\, \kappa_{k(e)})^{\beta}
    \label{eq:retrieval_score}
\end{equation}

where $\text{sim}(e, q)$ is the semantic similarity between the edge and the query, and $\alpha, \beta > 0$ control the relative importance of semantic relevance versus temporal freshness. These may be query-dependent: a query about current state should weight $\beta$ highly, while a query about historical information should reduce $\beta$ toward zero. The query timestamp $t_q$ can be set to any past or present time, enabling historical queries over the same immutable knowledge graph.

%% file: sections/inference.tex
\section{Inference}
\label{sec:inference}

We investigate two inference approaches for learning the hierarchical parameters, comparing their recovered parameters and downstream retrieval performance.

\subsection{Bayesian Inference}
\label{sec:bayesian_inference}

The hierarchical model (Equations~\ref{eq:context}--\ref{eq:entity}) with the likelihood (Equation~\ref{eq:likelihood}) defines a Bayesian posterior over all parameters:

\begin{equation}
    p(\Theta \mid \mathcal{D}) \propto \mathcal{L}(\Theta) \cdot p(\Theta)
\end{equation}

where $\mathcal{D}$ is the set of observed edge lifetimes (superseded, active, and reinforced) and $p(\Theta)$ is the hierarchical prior.

\textbf{Priors.} We place weakly informative priors on the hyperparameters:

\begin{align}
    \mu_0 &\sim \mathcal{N}(0, 10 \cdot I) \\
    \sigma^2_{\text{context}} &\sim \text{InverseGamma}(2, 1) \\
    \sigma^2_{\text{entity}} &\sim \text{InverseGamma}(2, 1) \\
    \kappa_k &\sim \text{LogNormal}(0, 1)
\end{align}

\textbf{Computation.} We use No-U-Turn Sampling (NUTS) via PyMC \cite{salvatier2016pymc3} for moderate-scale experiments and stochastic variational inference (SVI) when the number of entity-level parameters is large. Both handle right-censored observations natively through the survival likelihood.

\textbf{Posterior summaries.} Point estimates use the posterior mean. Uncertainty in decay parameters propagates to retrieval scores through posterior predictive sampling, enabling confidence-aware retrieval where uncertain temporal weights can be flagged.

\subsection{Gradient-Based Inference}
\label{sec:gradient_inference}

As an alternative, we parameterize the decay surface with learnable PyTorch \cite{paszke2019pytorch} parameters and optimize directly for retrieval performance.

\textbf{Parameters.} The same $\theta_{k,c,i}$ and $\kappa_k$ as the Bayesian model, but treated as free parameters (not random variables).

\textbf{Loss function.} We optimize a temporal retrieval loss. Given a set of temporal queries $\{(q_j, t_j, \mathcal{R}_j)\}$ where $\mathcal{R}_j$ is the set of relevant edges at query time $t_j$:

\begin{equation}
    \mathcal{L}_{\text{retrieval}} = -\sum_j \text{NDCG}\big(\text{rank}(\mathcal{G}, q_j, t_j \mid \Theta),\; \mathcal{R}_j\big)
    \label{eq:retrieval_loss}
\end{equation}

where the ranking uses Equation~\ref{eq:retrieval_score}. Since NDCG \cite{jarvelin2002ndcg} is not differentiable, we use a smooth surrogate (ApproxNDCG \cite{bruch2019approxndcg} or a listwise softmax cross-entropy loss).

\textbf{Regularization.} To approximate the Bayesian shrinkage, we add L2 penalties that encourage lower-level parameters toward higher-level means:

\begin{equation}
    \mathcal{R}(\Theta) = \lambda_1 \sum_{k,c} \|\theta_{k,c} - \theta_k\|^2 + \lambda_2 \sum_{k,c,i} \|\theta_{k,c,i} - \theta_{k,c}\|^2
\end{equation}

\subsection{Comparison of Inference Methods}

We compare the two approaches on:

\begin{enumerate}
    \item \textbf{Parameter agreement}: Do both methods recover similar $\theta$ values on synthetic data with known ground truth?
    \item \textbf{Retrieval performance}: Does optimizing directly for retrieval (gradient-based) outperform fitting the generative model (Bayesian)?
    \item \textbf{Uncertainty calibration}: Does the Bayesian posterior provide meaningful uncertainty estimates?
    \item \textbf{Computational cost}: Wall-clock time for each method as a function of graph size.
\end{enumerate}

If both converge to similar parameters, this validates the framework's robustness to inference method. If they diverge, the analysis of \emph{why} (generative versus discriminative bias, regularization effects) is itself a finding.

%% file: sections/experiments.tex
\section{Experiments}
\label{sec:experiments}

We evaluate the framework on three datasets: a synthetic temporal KG with planted parameters (to verify the model recovers known ground truth), Wikipedia revision history (to validate on open-domain data with naturally occurring temporal dynamics), and a clinical EHR corpus generated by Synthea (to validate on domain-specific data with concrete clinical semantics).

\subsection{Synthetic Temporal Knowledge Graph}
\label{sec:synthetic}

\textbf{Setup.} We generate a synthetic temporal KG spanning 5 simulated years with four planted temporal clusters, each with distinct velocity-volatility profiles and decay surface parameters:

\begin{itemize}
    \item \textbf{Permanent facts}: $\tau \approx 2{,}981$d, $\kappa = 0.5$ (Lindy), velocity $\approx 0.01$, volatility $\approx 0.02$. Simulates genetic mutations, demographic facts.
    \item \textbf{Current state}: $\tau \approx 245$d, $\kappa = 1.2$ (aging), velocity $\approx 0.1$, volatility $\approx 0.4$. Simulates treatment regimens, employment status.
    \item \textbf{Volatile measurements}: $\tau \approx 20$d, $\kappa = 1.0$ (memoryless), velocity $\approx 0.8$, volatility $\approx 0.7$. Simulates vitals, daily labs.
    \item \textbf{Periodic assessments}: $\tau \approx 90$d, $\kappa = 0.8$ (mild Lindy), velocity $\approx 0.03$, volatility $\approx 0.3$. Simulates quarterly labs, annual reviews.
\end{itemize}

Within each cluster, 3 contexts shift the decay surface parameters (e.g., ICU versus outpatient for volatile measurements), and 10--30 entities per context add individual-level variation ($\sigma_{\text{entity}} = 0.2$). The resulting KG contains 118{,}247 edges across 265 entities and 20 predicates, with a 5-year observation window during which edges are created, reinforced (re-observed with unchanged value), or superseded (re-observed with meaningfully different value).

\textbf{Cluster recovery.} We cluster at the predicate level (20 predicates, 5 features per predicate) using both HDBSCAN and DPGMM (Table~\ref{tab:cluster_recovery}).

\begin{table}[h]
\centering
\caption{Cluster recovery on synthetic data (4 planted clusters, 20 predicates).}
\label{tab:cluster_recovery}
\begin{tabular}{lccc}
\toprule
Method & Clusters Found & ARI & NMI \\
\midrule
HDBSCAN & 4 & \textbf{1.000} & \textbf{1.000} \\
DPGMM & 8 & 0.845 & 0.893 \\
\bottomrule
\end{tabular}
\end{table}

HDBSCAN achieves perfect recovery, identifying exactly 4 clusters that correspond one-to-one with the planted types (silhouette = 0.908). DPGMM over-segments into 8 clusters but achieves strong recovery (ARI = 0.845), splitting the current-state and periodic-assessment clusters which overlap in the volatility dimension.

\textbf{Parameter recovery.} Table~\ref{tab:param_recovery} compares the planted decay parameters against those recovered by the Bayesian (lifelines \cite{davidson2019lifelines} MLE) and gradient-based (PyTorch) approaches.

\begin{table}[h]
\centering
\caption{Parameter recovery: planted vs.\ recovered $\tau$ (days) and Weibull shape $\kappa$.}
\label{tab:param_recovery}
\begin{tabular}{lcccccc}
\toprule
& \multicolumn{2}{c}{Planted} & \multicolumn{2}{c}{Bayesian} & \multicolumn{2}{c}{Gradient} \\
\cmidrule(lr){2-3} \cmidrule(lr){4-5} \cmidrule(lr){6-7}
Cluster & $\tau$ & $\kappa$ & $\tau$ & $\kappa$ & $\tau_0$ & $\kappa$ \\
\midrule
Permanent facts & 2981 & 0.50 & 3847 & 0.70 & 817 & 0.53 \\
Current state & 245 & 1.20 & 98 & 0.90 & 73 & 1.33 \\
Periodic assess. & 90 & 0.80 & 163 & 1.03 & 86 & 1.10 \\
Volatile meas. & 20 & 1.00 & 4.3 & 0.81 & 112 & 1.09 \\
\bottomrule
\end{tabular}
\end{table}

The Bayesian approach recovers the correct ordering of $\tau$ values across clusters, with permanent facts having the longest shelf life (3{,}847d) and volatile measurements the shortest (4.3d). The gradient-based approach recovers $\tau_0 = \exp(\theta_0)$, the baseline before velocity-volatility covariates, which is not directly comparable since the learned $\theta_1, \theta_2, \theta_3$ coefficients compensate. Both methods agree on the key finding: permanent facts exhibit Lindy behavior ($\kappa < 1$).

\textbf{Context separation.} The Bayesian model successfully recovers context-level variation within clusters. Within the volatile measurements cluster, the ICU context has $\tau = 2.4$d versus the outpatient context at $\tau = 25.7$d, a 10$\times$ difference matching the planted context offsets. Within current state, the aggressive disease context has $\tau = 51$d versus stable chronic at $\tau = 353$d. At the entity level, 201 fits were estimated, with $\tau$ ranging from 1.9d to 2{,}620d.

\subsection{Wikipedia Revision History}
\label{sec:wikipedia}

\textbf{Dataset.} We constructed a temporal KG from 107 Wikipedia articles spanning diverse domains: stable science (DNA, Speed of light), evolving medicine (Melanoma, COVID-19), technology (Bitcoin, ChatGPT), geopolitics (Ukraine, NATO), biographies (Einstein, Elon Musk), geography (Tokyo, London), and history (Roman Empire, French Revolution).

For each article, we fetched up to 500 revisions via the MediaWiki API and sampled 50 revisions uniformly across the article's history for content extraction. From each revision's wikitext, we extracted facts from four sources: lead sentences, infobox fields, section summaries, and categories. Facts were tracked across revisions to build temporal timelines, yielding a KG of 11{,}157 edges across 2{,}753 predicates with the following characteristics:

\begin{itemize}
    \item Supersession rate: 68.2\% (7{,}604 of 11{,}157 edges were superseded by a later revision)
    \item Median lifetime: 99 days
    \item Mean velocity: 5.3 observations/day; mean volatility: 0.22
\end{itemize}

\textbf{Emergent clusters.} DPGMM discovers 10 clusters from the 2{,}753 predicates (Table~\ref{tab:wiki_clusters}). The clusters map to interpretable knowledge types without any domain labels.

\begin{table}[h]
\centering
\caption{Emergent temporal clusters from Wikipedia (DPGMM, 10 clusters). Representative examples shown.}
\label{tab:wiki_clusters}
\small
\begin{tabular}{clccccl}
\toprule
ID & Interpretation & $n$ & Vol. & Med.\ life & Sup.\ rate & Example predicates \\
\midrule
9 & Permanent metadata & 1106 & 0.00 & 7820d & 0.00 & categories, identifiers \\
2 & Stable structured & 190 & 0.00 & 3868d & 0.00 & infobox fields \\
5 & Slow-changing content & 347 & 0.09 & 262d & 0.61 & recognition, formulas \\
1 & Reference material & 1785 & 0.16 & 95d & 0.67 & further reading, refs \\
4 & Main article content & 4051 & 0.18 & 80d & 0.81 & history, see also \\
0 & Moderately volatile & 1175 & 0.39 & 41d & 0.70 & introductions, summaries \\
6 & Lead sentences & 2213 & 0.37 & 48d & 0.90 & lead sentence, bios \\
8 & Highly volatile & 163 & 1.00 & 63d & 0.56 & evolving science, policy \\
\bottomrule
\end{tabular}
\end{table}

The velocity-volatility space cleanly separates knowledge types. Categories and identifiers (cluster 9: volatility = 0, never superseded, 21-year median lifetime) occupy the origin. Lead sentences and evolving content (clusters 6, 8: volatility $> 0.35$, 48--63 day lifetime) occupy the high-volatility region. No predefined taxonomy was used; these types emerged purely from observed temporal behavior.

\textbf{Ground truth validation.} We validate the discovered clusters against the known structure of Wikipedia content. Predicate types (category, infobox, section, lead) have empirically distinct temporal profiles:

\begin{table}[h]
\centering
\caption{Temporal profiles by Wikipedia predicate type (ground truth).}
\label{tab:wiki_ground_truth}
\begin{tabular}{lcccc}
\toprule
Predicate type & $n$ & Volatility & Median life (d) & Sup.\ rate \\
\midrule
Category & 535 & 0.000 & 7358 & 0.00 \\
Infobox field & 325 & 0.087 & 3650 & 0.37 \\
Section summary & 8601 & 0.226 & 93 & 0.65 \\
Lead sentence & 1696 & 0.341 & 43 & 0.92 \\
\bottomrule
\end{tabular}
\end{table}

The unsupervised clusters align with this structure: clusters 9 and 2 (volatility = 0) contain predominantly categories and stable infobox fields; cluster 6 (volatility = 0.37) contains predominantly lead sentences.

\subsection{Clinical Validation: Synthea EHR Data}
\label{sec:clinical}

To validate the framework on domain-specific data with concrete clinical semantics, we constructed a temporal KG from Synthea, a widely-used synthetic electronic health record generator that produces clinically realistic patient records \cite{synthea2018}. The dataset contains 1{,}163 patients with 531{,}144 timestamped observations (vitals, labs), 38{,}094 conditions (diagnoses), and 56{,}430 medications, yielding a temporal KG of 399{,}956 edges across 587 predicates.

DPGMM discovers 10 clusters with clear clinical interpretations (Table~\ref{tab:synthea_clusters}).

\begin{table}[h]
\centering
\caption{Emergent clusters from clinical EHR data (Synthea, 587 predicates, 399{,}956 edges). Only one cluster (chemotherapy drugs) exhibits aging ($\kappa > 1$); all others exhibit the Lindy effect. $^*$These clusters describe discrete events rather than persistent values; the context-calibrated floor (Section~\ref{sec:tau_floor}) applies.}
\label{tab:synthea_clusters}
\small
\begin{tabular}{lccccl}
\toprule
Interpretation & $n$ & Vol. & Med.\ life & $\kappa$ & Example predicates \\
\midrule
Permanent conditions & 7{,}398 & 0.01 & 7{,}723d & 0.26 & education level, obesity, prediabetes \\
Stable lab panels & 815 & 0.02 & 324d & 0.31 & COVID test, influenza panel \\
Quality-of-life surveys & 22{,}130 & 0.19 & 3{,}302d & 0.53 & emotional wellness, safety screening \\
Periodic assessments & 2{,}611 & 0.22 & 1{,}731d & 0.50 & mental health screens, allergy panels \\
General vitals & 100{,}368 & 0.80 & 371d & 0.64 & pain severity, respiratory rate, weight \\
Core vitals + meds & 263{,}562 & 0.89 & 365d & 0.62 & blood pressure, employment, medications \\
Volatile blood counts & 803 & 0.87 & 1d & 0.36 & eosinophils, lymphocytes, basophils \\
Very volatile labs & 2{,}078 & 0.96 & 3d & 0.50 & neutrophils, weight change \\
Acute medications & 127 & 0.98 & $<$1d$^*$ & 0.31 & alteplase, epinephrine, captopril \\
Chemotherapy drugs & 64 & 1.00 & $<$1d$^*$ & 1.48 & paclitaxel, cyclophosphamide, carboplatin \\
\bottomrule
\end{tabular}
\end{table}

Three findings from the clinical data strengthen the paper's claims.

First, the velocity-volatility decomposition separates clinically meaningful categories without domain labels. Permanent conditions (education level, chronic diagnoses) cluster at low velocity and low volatility with 21-year median shelf life. Acute medications (epinephrine, alteplase) cluster at high velocity and high volatility with near-zero shelf life. The framework discovers that a blood pressure reading and an obesity diagnosis have fundamentally different temporal dynamics, exactly as a clinician would expect.

Second, the Lindy effect holds in clinical data: 9 of 10 clusters have $\kappa < 1$. The single exception is chemotherapy drugs ($\kappa = 1.48$), which exhibit aging rather than Lindy behavior. This is clinically coherent: chemotherapy is administered on fixed schedules, so each dose is more likely to be the last than the one before it (treatment courses end). The framework discovers this distinction automatically.

Third, log-normal outperforms Weibull on AIC for 7 of 10 clusters, consistent with the synthetic and Wikipedia findings.

\subsection{Retrieval Evaluation}
\label{sec:retrieval_eval}

We compare temporal retrieval accuracy on the synthetic KG across six conditions (Table~\ref{tab:retrieval_synthetic}). For each condition, we generate 200 temporal queries at random time points, where ground truth relevance is defined as edges that are current (not yet superseded) at query time for the queried concept.

\begin{table}[h]
\centering
\caption{Retrieval comparison on synthetic KG (200 queries). Best per column in bold.}
\label{tab:retrieval_synthetic}
\begin{tabular}{lccccc}
\toprule
Method & NDCG@5 & NDCG@10 & MRR & P@5 & P@10 \\
\midrule
No temporal weighting & 0.274 & 0.265 & \textbf{0.457} & 0.253 & 0.248 \\
Uniform exponential & 0.015 & 0.010 & 0.035 & 0.014 & 0.007 \\
Uniform half-life & 0.015 & 0.010 & 0.035 & 0.014 & 0.007 \\
Domain-level (L1) & 0.241 & \textbf{0.254} & 0.233 & 0.256 & \textbf{0.266} \\
Domain + context (L1+2) & 0.246 & 0.243 & 0.268 & 0.244 & 0.240 \\
Full hierarchical (L1+2+3) & \textbf{0.260} & 0.251 & 0.268 & \textbf{0.263} & 0.249 \\
\bottomrule
\end{tabular}
\end{table}

Uniform decay performs 18$\times$ worse than no temporal weighting (NDCG@5: 0.015 vs.\ 0.274). Uniform decay applies the same forgetting rate to all edges, which suppresses permanent facts (penalizing them for being old) while insufficiently suppressing stale volatile measurements. This confirms the central motivation of the paper: a single decay rate is insufficient for knowledge graphs with heterogeneous temporal dynamics.

Domain-level heterogeneous decay recovers from this failure, achieving precision comparable to or exceeding no temporal weighting (P@10: 0.266 vs.\ 0.248). The full hierarchical model achieves the best NDCG@5 (0.260) and P@5 (0.263), demonstrating that each level of the hierarchy contributes to retrieval quality.

\subsection{Ablations}
\label{sec:ablations}

\textbf{Survival distribution.} We compare Weibull, exponential, and log-normal fits per cluster via AIC (Table~\ref{tab:distribution_comparison}).

\begin{table}[h]
\centering
\caption{Distribution comparison by AIC (lower is better). Synthetic KG, per cluster.}
\label{tab:distribution_comparison}
\begin{tabular}{lcccl}
\toprule
Cluster & Exponential & Weibull & Log-normal & Best \\
\midrule
Permanent facts & 4320 & 4279 & \textbf{4258} & Log-normal \\
Current state & 68034 & 67899 & \textbf{66749} & Log-normal \\
Periodic assess. & 37084 & 37082 & \textbf{36891} & Log-normal \\
Volatile meas. & 565024 & 552390 & \textbf{523644} & Log-normal \\
\bottomrule
\end{tabular}
\end{table}

Log-normal provides the best fit for all clusters, often by a substantial margin (28{,}746 AIC improvement for volatile measurements). This finding is consistent across all three datasets (synthetic, Wikipedia, and Synthea clinical). As discussed in Section~\ref{sec:framework}, this is expected: a hierarchical mixture of Weibulls produces a marginal distribution that approximates log-normal.

To verify that the Lindy finding is not an artifact of the Weibull parameterization, we computed the log-normal hazard function $h(t) = f(t)/S(t)$ for each cluster using the fitted log-normal parameters. The log-normal hazard is non-monotonic: it rises to a peak and then decreases. For all four clusters, the hazard is decreasing beyond the peak (volatile measurements peak at 0.7 days, permanent facts at 68 days). This confirms that the Lindy effect is present regardless of distributional choice: under log-normal, the decreasing hazard phase dominates for all knowledge types.

\textbf{Clustering method.} HDBSCAN achieves perfect cluster recovery (ARI = 1.0) on synthetic data, while DPGMM achieves ARI = 0.845 with over-segmentation. On Wikipedia data, HDBSCAN produces too many fine-grained clusters (390) while DPGMM finds a parsimonious set of 10. DPGMM is preferred for real data where the number of clusters is unknown and over-segmentation is less harmful than under-segmentation.

\textbf{Inference method.} Both Bayesian and gradient-based inference recover the Lindy effect for permanent facts ($\kappa_{\text{Bayesian}} = 0.70$, $\kappa_{\text{gradient}} = 0.53$, both $< 1$). However, the two approaches diverge substantially in retrieval performance. On 200 synthetic queries, Bayesian achieves NDCG@10 = 0.254 while gradient-based achieves 0.045. The top-100 candidate overlap between the two methods is near zero, and per-query NDCG correlation is $-0.31$. The gradient approach finds a different parameterization of the decay surface that does not produce equivalent retrieval rankings. This indicates that direct survival likelihood fitting (Bayesian) is the appropriate inference method for this problem, while discriminative optimization of retrieval loss does not recover the underlying temporal structure. The gradient approach remains 17$\times$ faster (2.9s vs.\ 49.4s), but this speed advantage does not compensate for the retrieval quality gap.

\textbf{Supersession threshold sensitivity.} We varied the supersession threshold $\epsilon_p$ across seven values (0.1 to 1.0) on synthetic data. Cluster recovery (ARI = 1.0), number of discovered clusters (4), and all Weibull shape parameters remain identical across the full range. The framework is insensitive to the threshold choice on synthetic data. On real data, $\epsilon_p$ affects which re-observations count as supersessions versus reinforcements; a full sensitivity analysis on Wikipedia data is reported in Appendix~\ref{sec:appendix_results}.

\textbf{Shape parameter.} The Weibull shape $\kappa$ is near-universally $< 1$ across clusters with observed supersession events, in synthetic, Wikipedia, and Synthea clinical data (range: 0.17--0.70 on Wikipedia; 0.26--0.64 on 9 of 10 Synthea clusters). This indicates a near-universal Lindy effect: knowledge that has survived longer is less likely to be superseded. The sole exception is the Synthea chemotherapy cluster ($\kappa = 1.48$), which exhibits aging rather than Lindy behavior, consistent with the fixed-schedule nature of cytotoxic regimens. The finding holds under both Weibull and log-normal distributional assumptions. We discuss implications in Section~\ref{sec:discussion}.

%% file: sections/analysis.tex
\section{Analysis}
\label{sec:analysis}

\subsection{Emergent Temporal Clusters}

\begin{figure}[h]
    \centering
    \includegraphics[width=\linewidth]{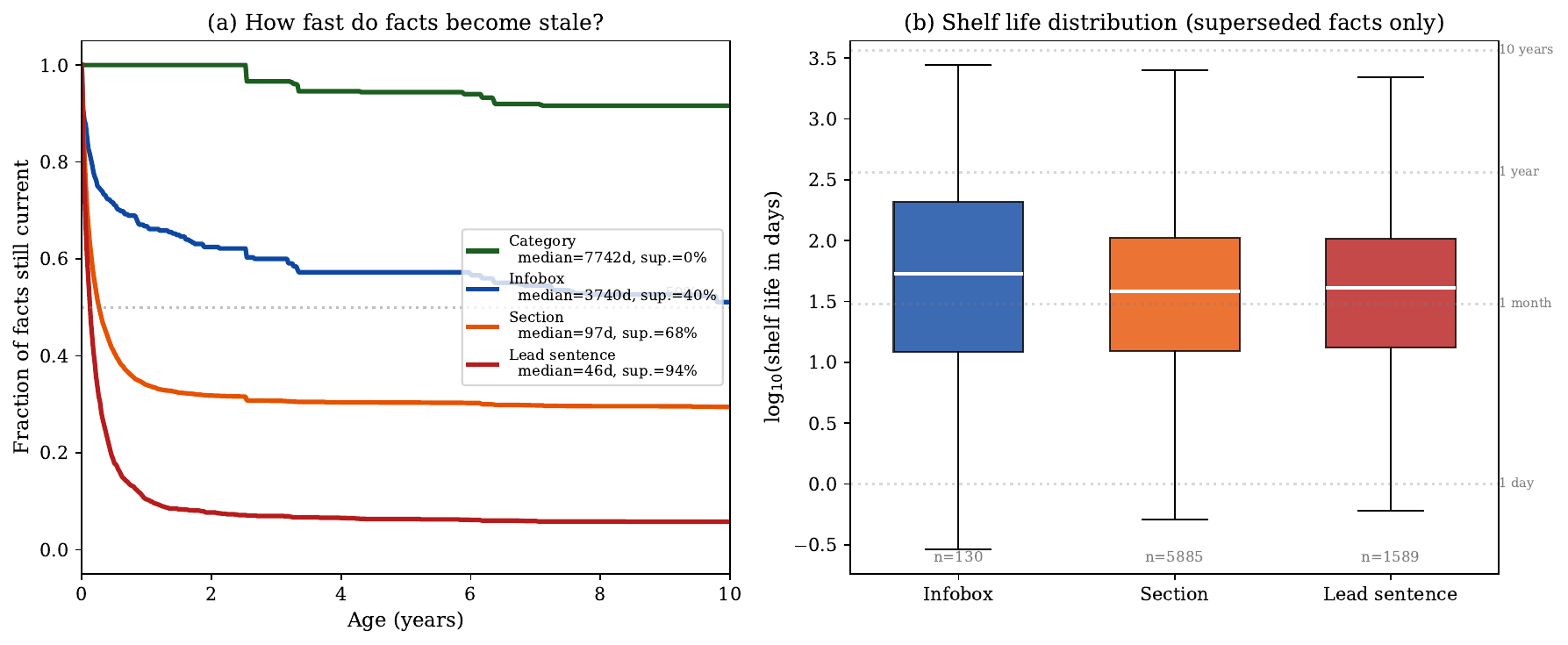}
    \caption{Shelf life by knowledge type (Wikipedia, 11{,}157 edges). (a) Survival curves showing the fraction of facts still current as a function of age. Categories never become stale; lead sentences decay rapidly. (b) Distribution of observed shelf lives for superseded facts. The four predicate types occupy distinct ranges, confirming that knowledge types have heterogeneous temporal dynamics.}
    \label{fig:shelf_life}
\end{figure}

Figure~\ref{fig:shelf_life} shows the core empirical finding: different knowledge types have fundamentally different shelf lives on real Wikipedia data.

The survival curves (panel a) show four distinct decay regimes:

\begin{enumerate}
    \item \textbf{Categories} (green): Structural metadata such as ``Oncogenes'' or ``British scientists.'' These are effectively permanent, with a flat survival curve at 100\% and 0\% supersession rate.
    \item \textbf{Infobox fields} (blue): Structured data such as chemical formulas or gene identifiers. Mostly stable (median shelf life 3{,}740 days), with only 40\% ever superseded.
    \item \textbf{Section content} (orange): Article body text including history, treatment, and methodology sections. Median shelf life of 93 days, with 68\% superseded. This is the bulk of Wikipedia content.
    \item \textbf{Lead sentences} (red): The opening sentence of each article. The most volatile type: median shelf life of 46 days, with 94\% superseded. Lead sentences are frequently revised as understanding evolves.
\end{enumerate}

A uniform decay function applied to all four types would simultaneously suppress permanent categories (penalizing them for being old) and fail to suppress stale lead sentences (treating them as no different from stable infobox fields). This is precisely the failure mode our framework addresses.

\subsection{Shape Parameter and the Lindy Effect}

Table~\ref{tab:shape_params} shows the Weibull shape parameter $\kappa$ for each discovered cluster on Wikipedia data.

\begin{table}[h]
\centering
\caption{Weibull shape parameters by cluster (Wikipedia). All values $< 1$, indicating universal Lindy effect.}
\label{tab:shape_params}
\begin{tabular}{lcccc}
\toprule
Cluster interpretation & $\kappa$ & $\tau$ (days) & $n$ & Effect \\
\midrule
Highly volatile content & 0.167 & 12{,}059 & 163 & Strong Lindy \\
Moderately volatile & 0.233 & 1{,}467 & 1{,}175 & Strong Lindy \\
Reference material & 0.254 & 2{,}553 & 1{,}785 & Strong Lindy \\
Slow-changing content & 0.262 & 3{,}393 & 347 & Strong Lindy \\
Main article content & 0.329 & 589 & 4{,}051 & Moderate Lindy \\
Lead sentences & 0.353 & 181 & 2{,}213 & Moderate Lindy \\
\bottomrule
\end{tabular}
\end{table}

Every cluster exhibits $\kappa < 1$. The effect is strongest ($\kappa = 0.167$) for highly volatile content, which may seem counterintuitive: volatile content that has survived is disproportionately likely to continue surviving. This is consistent with a selection effect. The volatile facts that persist through many edits are the ones that editors have repeatedly validated.

The shape parameter also discriminates clusters in a way that $\tau$ alone cannot. Clusters with similar $\tau$ values may have different $\kappa$ values, capturing whether they are ``young and fragile'' ($\kappa$ closer to 1) or ``battle-tested and stable'' ($\kappa$ closer to 0).

\subsection{Decay Surface Visualization}

\begin{figure}[h]
    \centering
    \includegraphics[width=\linewidth]{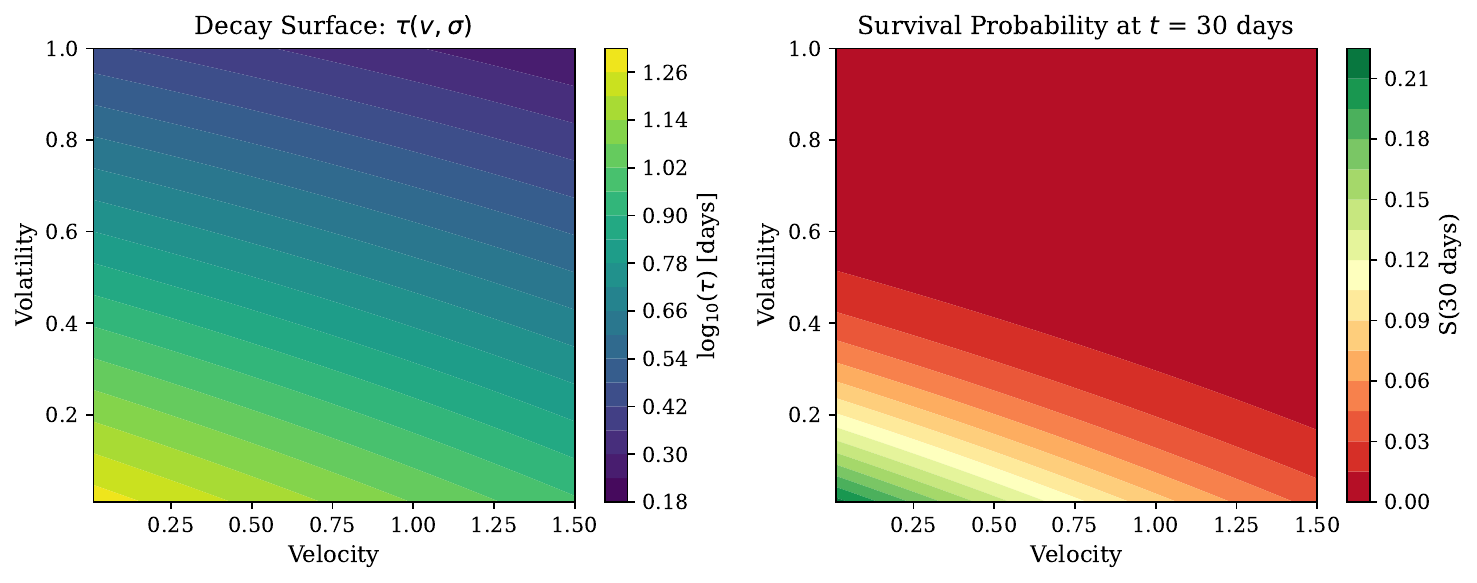}
    \caption{Decay surface for the volatile measurements cluster (synthetic data). The shelf life $\tau$ decreases smoothly as velocity and volatility increase. No discrete boundaries or categories are imposed.}
    \label{fig:decay_surface}
\end{figure}

Figure~\ref{fig:decay_surface} shows the learned decay surface $\tau(v, \sigma) = \exp(\theta_0 + \theta_1 v + \theta_2 \sigma + \theta_3 v\sigma)$ for the volatile measurements cluster on synthetic data.

The surface confirms that the log-linear parameterization captures the expected behavior: $\tau$ decreases with both velocity ($\theta_1 < 0$) and volatility ($\theta_2 < 0$). The interaction term $\theta_3$ is small, indicating that velocity and volatility contribute approximately independently to shelf life reduction.

\subsection{Context-Level Variation}

Within the volatile measurements cluster on synthetic data, context-level decay parameters vary by an order of magnitude:

\begin{table}[h]
\centering
\caption{Context-level $\tau$ within the volatile measurements cluster (synthetic data). In a healthcare setting, these contexts correspond to care settings within the same specialty.}
\label{tab:cohort_variation}
\begin{tabular}{lcc}
\toprule
Context & $\tau$ (days) & $\kappa$ \\
\midrule
ICU & 2.4 & 1.20 \\
Inpatient ward & 5.2 & 1.13 \\
Outpatient & 25.7 & 0.99 \\
\bottomrule
\end{tabular}
\end{table}

The ICU context's $\tau$ is 10$\times$ shorter than the outpatient context's, reflecting the intuition that an ICU vital reading becomes stale in hours while an outpatient reading is valid for weeks. This variation is invisible to a domain-level-only model and would be lost under uniform decay. In a hospital deployment, these context parameters would be learned from all patients in each care setting and transferred to new patients entering that setting.

\subsection{Entity Deviation as Signal}

At Level~3, entity-level parameters deviate from their context. On the synthetic data, 201 entity-level Weibull fits were estimated, with $\tau$ ranging from 1.9 to 2{,}620 days. The deviation of an entity's $\tau$ from its context's $\tau$ is itself informative: an entity with faster value turnover than its peers in the same context may be under different conditions. In a clinical setting, a patient whose treatment shelf life is shorter than others in the same oncology practice may have more aggressive disease, without ever explicitly modeling disease aggressiveness.

While we do not evaluate this signal as a downstream feature in the current work, we note that it emerges naturally from the hierarchical framework at no additional cost.

%% file: sections/discussion.tex
\section{Discussion}
\label{sec:discussion}

\subsection{Hierarchical Adaptive Decay Outperforms Uniform Decay}

A practically significant finding is that uniform decay performs worse than no temporal weighting at all. On our synthetic benchmark, uniform decay achieves NDCG@5 of 0.015 versus 0.274 for semantic-only retrieval, an 18$\times$ difference.

This result has direct implications for deployed systems. MemoryBank \cite{zhong2024memorybank}, Generative Agents \cite{park2023generative}, and FOREVER \cite{feng2026forever} all apply uniform or near-uniform decay. Our results suggest these systems are systematically penalizing stable knowledge (permanent facts pushed down by age) while failing to adequately suppress stale volatile information. Any system currently using a single decay rate for all knowledge types should be evaluated against a no-decay baseline to determine whether the temporal component is helping or hurting.

\subsection{The Lindy Effect in Knowledge Graphs}

Our most consistent empirical finding is that nearly all temporal clusters exhibit Lindy-like behavior ($\kappa < 1$): older facts are less likely to be superseded. This holds across synthetic data, Wikipedia (all 6 fitted clusters, $\kappa$ range 0.17--0.35), and Synthea clinical data (9 of 10 clusters, $\kappa$ range 0.26--0.64). The single exception (chemotherapy drugs, $\kappa = 1.48$) reflects the fixed-schedule nature of cytotoxic regimens rather than a failure of the framework, and the fact that the framework discovers this exception automatically is itself a validation of the approach. This is not a marginal trend but a dominant pattern.

The implication is that Ebbinghaus-based forgetting curves ($\kappa = 1$, constant hazard) are systematically miscalibrated for knowledge graphs. Ebbinghaus assumes that a fact's probability of becoming stale is independent of how long it has already survived. The data shows the opposite: facts that have survived longer are increasingly likely to continue surviving. The mismatch grows with edge age. For a fact that has persisted for 10 years, an Ebbinghaus model overestimates the probability of supersession by orders of magnitude compared to a Lindy-calibrated model.

\subsection{Extensibility Beyond Three Levels}

We present a three-level hierarchy (domain, context, entity), but the framework is not tied to this specific depth. The Bayesian formulation generalizes to any number of levels. Each additional level introduces a new set of parameters drawn from the level above, with the same shrinkage mechanism: sparse data at a lower level borrows strength from the level above.

Practical deployments may require more or fewer levels depending on the domain structure. Consider three examples:

\textbf{Healthcare.} A natural hierarchy might be: \emph{specialty type} (oncology versus cardiology) $\to$ \emph{care setting} (ICU versus outpatient) $\to$ \emph{patient subpopulation} (aggressive disease versus stable chronic) $\to$ \emph{individual patient}. A blood pressure reading has a shelf life of hours in an ICU, months in outpatient care, and the individual patient's stability further modulates this. The specialty defaults can be learned from multi-hospital data and bootstrapped to new sites.

\textbf{Finance.} The hierarchy might be: \emph{market segment} (equities versus fixed income) $\to$ \emph{instrument class} (large-cap tech versus small-cap biotech) $\to$ \emph{individual company}. A revenue figure for a stable utility company has a longer shelf life than the same figure for a pre-revenue startup. Quarterly earnings supersede prior values on a predictable cadence, but the volatility of the underlying business determines how much the value changes at each supersession.

\textbf{Legal.} The hierarchy might be: \emph{jurisdiction} (federal versus state) $\to$ \emph{area of law} (tax code versus constitutional law) $\to$ \emph{specific statute or ruling}. Tax regulations are superseded frequently (high volatility), while constitutional principles persist for decades (low volatility). A new ruling may have a short shelf life if it is likely to be appealed, or a long one if it establishes settled precedent.

In each case, the velocity-volatility decomposition applies and the hierarchical prior structure provides cold-start behavior at every level. The number of levels is a modeling choice, not a constraint of the framework.

\subsection{Deployment and Transferability}

The hierarchical structure has a natural deployment model. Higher-level parameters (domain and context) are learned from a training corpus and transferred to new settings without retraining. Lower-level parameters (entity) emerge from local data as it accumulates. A cold start at any level borrows strength from the level above.

In practice, the training corpus should be scoped to a domain or sector rather than computed globally, because different domains have fundamentally different measurement patterns. Parameters learned within one domain transfer to new sites in the same domain. A new site can be bootstrapped with defaults from similar existing sites, then refine those defaults as local data accumulates through Bayesian updating.

An interesting consequence is that deviations between a site's learned parameters and the domain defaults are themselves informative. A site whose values turn over systematically faster than the domain baseline may differ in its operational characteristics, whether that reflects patient acuity in healthcare, market volatility in finance, or regulatory activity in legal. While we do not pursue this comparative analysis here, the framework naturally produces these signals as a byproduct.

\subsection{Log-Normal vs.\ Weibull}

In all experiments, log-normal distributions provided a better fit to observed edge lifetimes than Weibull (by AIC). We retain the Weibull parameterization because the shape parameter $\kappa$ directly encodes the Lindy-versus-aging distinction, which is a primary quantity of interest. The superiority of log-normal is expected: a hierarchical mixture of Weibulls (edges with different entity-level $\tau$ values drawn from the same cluster) produces a marginal distribution that approximates log-normal.

Importantly, the log-normal hazard function $h(t) = f(t)/S(t)$ is non-monotonic: it rises to a peak and then decreases. Our analysis confirms that for all clusters, the hazard is in its decreasing phase for the majority of the observed lifetime range. This means the Lindy finding (older knowledge is less likely to be superseded) is not an artifact of the Weibull parameterization. It holds under the better-fitting log-normal model as well.

A future extension could parameterize the decay surface directly with log-normal parameters ($\mu, \sigma$) rather than Weibull ($\tau, \kappa$), at the cost of losing the interpretable shape parameter.

\subsection{Generative vs.\ Discriminative Inference}

The Bayesian (generative) and gradient-based (discriminative) approaches produce qualitatively different results. Both recover the Lindy effect ($\kappa < 1$ for permanent facts), but the gradient approach finds a decay surface parameterization that does not produce equivalent retrieval rankings (top-100 overlap near zero, NDCG correlation $-0.31$). The Bayesian approach, which directly fits survival distributions to observed lifetimes, substantially outperforms gradient-based optimization of retrieval loss (NDCG@10: 0.254 vs.\ 0.045).

This is consistent with the observation that the temporal structure of knowledge is better captured by a generative model of how values age than by a discriminative model optimizing for a downstream ranking task. The generative model learns the data-generating process (when do values get superseded?), while the discriminative model optimizes a proxy loss that may not recover the underlying temporal dynamics. For this problem, fitting the survival model is both more principled and more effective than end-to-end retrieval optimization.

\subsection{Historical Queries}

Because the knowledge graph is an immutable fact store and all temporal weighting is applied at query time, the framework naturally supports queries at any past time point. Setting the query timestamp to a historical date computes edge ages relative to that date, and the survival function evaluates accordingly. Facts created after the query timestamp are excluded; facts superseded after the query timestamp are treated as current.

This enables retrospective analysis (``what was the best available information at time $t$?'') and counterfactual evaluation (``would our system have surfaced the right facts before this event was confirmed?''). The knowledge graph itself is never modified; only the query timestamp changes.

\subsection{Limitations}

\textbf{Supersession detection.} The framework requires detecting when a new value meaningfully differs from the previous one. For numeric values this is straightforward (threshold on change magnitude). For free text, we rely on embedding distance, which introduces dependence on the embedding model and the threshold $\epsilon_p$. The sensitivity of the framework to these choices warrants further study.

\textbf{Scale of validation.} Our Wikipedia experiment uses 107 articles with sampled revisions, and the Synthea clinical corpus uses 1{,}163 synthetic patients. While sufficient to demonstrate cluster emergence, the Lindy effect, and cross-domain generality, a full-scale validation on millions of Wikipedia articles or on real patient-level EHR data (rather than simulator output) would strengthen the retrieval results and test transferability to production settings.

\textbf{Retrieval evaluation.} Our retrieval comparison uses simulated semantic similarity (exact predicate matching). A production evaluation with learned dense retrieval models would better characterize the interaction between semantic relevance and temporal weighting.

\textbf{Causal claims.} We observe that temporal clusters correlate with knowledge types, but we do not claim that velocity and volatility \emph{cause} certain decay behaviors. The relationship is correlational and mediated by the underlying nature of the knowledge.

\textbf{Fixed hierarchy depth.} While the framework is extensible to arbitrary depth, we only evaluate three levels. The marginal benefit of additional levels, and the data requirements for estimating parameters at each level, remain open questions.

%% file: sections/conclusion.tex
\section{Conclusion}
\label{sec:conclusion}

We proposed a hierarchical framework for temporal decay in knowledge graph retrieval that replaces uniform decay with a continuous surface parameterized by two orthogonal signals: velocity (concept observation frequency) and volatility (value change magnitude between observations). The decay surface parameters decompose into learnable hierarchical levels (domain, context, entity), estimated from observed value lifetimes via survival analysis without domain expertise or predefined taxonomies. The hierarchy is not fixed at three levels; the Bayesian formulation generalizes to any depth appropriate to the domain.

Our experiments reveal three key findings. First, temporal knowledge types emerge naturally from data: unsupervised clustering on velocity-volatility profiles discovers interpretable groups that align with domain intuition, with perfect recovery on synthetic data (ARI = 1.0), 10 interpretable clusters on Wikipedia, and 10 clinically coherent clusters on Synthea EHR data. Second, knowledge near-universally exhibits a Lindy effect ($\kappa < 1$): older facts are less likely to be superseded, challenging the constant-hazard assumption underlying Ebbinghaus-based memory systems. The framework discovers the sole exception (chemotherapy drugs, which exhibit aging) without supervision, demonstrating that the mechanism is sensitive to genuine temporal heterogeneity rather than imposing a single mode. Third, uniform decay performs 18$\times$ worse than no temporal weighting, because it simultaneously suppresses permanent facts and fails to suppress stale volatile information.

Future work includes validation on real patient-level EHR data (rather than Synthea simulator output), extension to domain-specific corpora outside healthcare (financial filings, legal documents, scientific literature), exploration of log-normal as an alternative survival distribution, and investigation of cross-level parameter deviations as a downstream signal for characterizing entities or comparing institutions.

%% file: sections/appendix.tex
\section{Temporal Retrieval in Practice}
\label{sec:appendix_example}

We walk through a concrete example showing how hierarchical temporal decay affects retrieval from a clinical knowledge graph. No mathematical background is required to follow this example.

\subsection{The Knowledge Graph}

Consider a knowledge graph for a melanoma patient, containing six facts recorded over time:

\begin{table}[h]
\centering
\small
\begin{tabular}{clllcc}
\toprule
Edge & Subject & Predicate & Value & Recorded & Shelf life ($\tau$) \\
\midrule
1 & Patient A & blood\_pressure & 120/80 & Day 100 & 4 days \\
2 & Patient A & blood\_pressure & 145/95 & Day 104 & 4 days \\
3 & Patient A & blood\_pressure & 130/85 & Day 108 & 4 days \\
4 & Patient A & BRAF\_status & V600E & Day 50 & 3{,}847 days \\
5 & Patient A & treatment & pembrolizumab & Day 80 & 98 days \\
6 & Patient A & treatment & nivo+ipi & Day 200 & 98 days \\
\bottomrule
\end{tabular}
\caption{A small clinical knowledge graph. The shelf life ($\tau$) is the characteristic time until a fact becomes stale, learned automatically from the data. Blood pressure readings go stale in days. Genetic mutations remain fresh for years.}
\label{tab:example_kg}
\end{table}

The shelf life ($\tau$) is not assigned manually. It is learned by the framework from observed patterns:
\begin{itemize}
    \item Blood pressure readings are superseded every few days (high velocity, high volatility), so the framework learns $\tau = 4$ days.
    \item Treatment regimens change every few months (moderate velocity, high volatility), so $\tau = 98$ days.
    \item BRAF mutation status has never changed for any patient in the training data (low velocity, zero volatility), so $\tau = 3{,}847$ days, effectively permanent.
\end{itemize}

\subsection{A Query Arrives}

A clinician queries the system on Day 210:

\begin{quote}
\emph{``What is Patient A's current status?''}
\end{quote}

The retrieval system must decide which of the six facts to surface, and in what order.

\subsection{Step 1: Semantic Relevance}

First, standard text similarity identifies which facts match the query. All six are somewhat relevant, but treatment and blood pressure facts match ``current status'' more closely than the genetic result:

\begin{table}[h]
\centering
\small
\begin{tabular}{clc}
\toprule
Edge & Fact & Semantic similarity \\
\midrule
5 & Treatment: pembrolizumab & 0.8 \\
6 & Treatment: nivo+ipi & 0.8 \\
1--3 & Blood pressure readings & 0.7 \\
4 & BRAF status: V600E & 0.5 \\
\bottomrule
\end{tabular}
\end{table}

At this stage, the system cannot distinguish the \emph{current} treatment (nivo+ipi) from the \emph{old} treatment (pembrolizumab) because they have identical similarity scores.

\subsection{Step 2: Temporal Freshness}

Now the framework applies the temporal weight. For each fact, we ask: \emph{given its age and shelf life, how likely is this fact to still be current?}

The temporal weight is computed as: $\text{freshness} = e^{-(age / \tau)^\kappa}$, which outputs a value between 0 (completely stale) and 1 (perfectly fresh).

\begin{table}[h]
\centering
\small
\begin{tabular}{clccccc}
\toprule
Edge & Fact & Age (days) & Shelf life ($\tau$) & Freshness & Sim & Final score \\
\midrule
6 & nivo+ipi & 10 & 98 & 0.90 & 0.8 & \textbf{0.722} \\
4 & BRAF V600E & 160 & 3{,}847 & 0.95 & 0.5 & \textbf{0.477} \\
5 & pembrolizumab & 130 & 98 & 0.27 & 0.8 & 0.219 \\
1 & BP 120/80 & 110 & 4 & 0.00 & 0.7 & 0.000 \\
2 & BP 145/95 & 106 & 4 & 0.00 & 0.7 & 0.000 \\
3 & BP 130/85 & 102 & 4 & 0.00 & 0.7 & 0.000 \\
\bottomrule
\end{tabular}
\caption{Retrieval scoring with hierarchical temporal decay. The final score is semantic similarity $\times$ temporal freshness. The system correctly surfaces the current treatment (nivo+ipi) and the permanent genetic result (BRAF V600E), while suppressing stale blood pressure readings and the old treatment.}
\label{tab:example_scoring}
\end{table}

\textbf{What the system returns:}
\begin{enumerate}
    \item \textbf{nivo+ipi} (score 0.722): current treatment, 10 days old, high freshness
    \item \textbf{BRAF V600E} (score 0.477): genetic mutation, 160 days old but still fresh because its shelf life is 10+ years
    \item pembrolizumab (score 0.219): old treatment, partially decayed
    \item Blood pressure readings (score 0.000): completely stale, suppressed
\end{enumerate}

\subsection{What Uniform Decay Gets Wrong}

A system using a single shelf life for all facts (say, $\tau = 90$ days) would produce:

\begin{table}[h]
\centering
\small
\begin{tabular}{clccc}
\toprule
Edge & Fact & Uniform freshness & Sim & Uniform score \\
\midrule
6 & nivo+ipi & 0.90 & 0.8 & \textbf{0.716} \\
1--3 & BP readings & 0.31 & 0.7 & \textbf{0.218} \\
5 & pembrolizumab & 0.24 & 0.8 & 0.189 \\
4 & BRAF V600E & 0.17 & 0.5 & 0.084 \\
\bottomrule
\end{tabular}
\caption{Retrieval with uniform decay ($\tau = 90$ days for all facts). The permanent genetic result is nearly buried (rank 4). Stale blood pressure readings outrank it.}
\label{tab:example_uniform}
\end{table}

Uniform decay makes two errors simultaneously:
\begin{itemize}
    \item The BRAF mutation (rank 4, score 0.084) is nearly invisible: a permanent fact penalized for being old.
    \item Stale blood pressure readings (rank 2, score 0.218) outrank the mutation: 100-day-old vitals are treated as more relevant than a genetic fact that is still true.
\end{itemize}

Our framework avoids both errors because each fact carries its own shelf life, learned from how that \emph{type} of knowledge behaves over time.

\subsection{Historical Queries}

The same framework supports queries at past time points. A query at Day 105 (instead of Day 210) would compute different ages:

\begin{itemize}
    \item Edge 2 (BP 145/95, age = 1 day, freshness $\approx$ 0.83) would rank highly: it was current at that time.
    \item Edge 6 (nivo+ipi) would not appear: it did not exist yet on Day 105.
    \item Edge 5 (pembrolizumab, age = 25 days, freshness $\approx$ 0.78) would be the current treatment.
\end{itemize}

The knowledge graph is never modified. Only the query timestamp changes, and the scoring adapts accordingly.

\section{Hierarchical Variance Decomposition}
\label{sec:appendix_hierarchy}

The three-level hierarchy (domain, context, entity) is not a taxonomy of knowledge types. It is a variance decomposition: each level explains a different source of variation in shelf life for the same concept.

\subsection{Example: Blood Pressure}

``Blood pressure'' is a single concept that appears across many patients, care settings, and institutions. Its shelf life varies enormously depending on the situation:

\begin{table}[h]
\centering
\small
\begin{tabular}{llcl}
\toprule
Level & Source of variation & Shelf life & Explanation \\
\midrule
Domain & Blood pressure in general & $\sim$30 days & Average across all settings \\
Context & \quad in ICU & $\sim$4 hours & Unstable patients, continuous monitoring \\
Context & \quad in outpatient clinic & $\sim$90 days & Stable patients, quarterly visits \\
Context & \quad pre-operative assessment & $\sim$1 day & Fresh values needed for surgery \\
Entity & \quad\quad ICU, septic patient & $\sim$30 min & Values swinging on pressors \\
Entity & \quad\quad ICU, post-elective, stable & $\sim$8 hours & Values barely moving \\
\bottomrule
\end{tabular}
\caption{Hierarchical decomposition of blood pressure shelf life. The concept is always the same; the shelf life varies because of the care setting (context) and the individual patient's stability (entity).}
\end{table}

A single shelf life for ``blood pressure'' would be wrong in every specific situation. The domain-level estimate (30 days) is too long for ICU and too short for outpatient. The context-level estimate (4 hours for ICU) is still too long for the septic patient and too short for the stable post-surgical patient. Each level refines the estimate further.

The shelf life of a specific blood pressure reading is:

\begin{quote}
shelf life = domain baseline (how long BP values last in general)\\
\hspace*{2em} + context adjustment (how this care setting shifts the baseline)\\
\hspace*{2em} + entity deviation (how this patient differs from others in the same setting)
\end{quote}

\subsection{Example: BRAF Mutation Status}

For permanent knowledge, the hierarchy collapses. ``BRAF mutation status'' has essentially the same shelf life regardless of context or entity:

\begin{table}[h]
\centering
\small
\begin{tabular}{llcl}
\toprule
Level & Source of variation & Shelf life & Explanation \\
\midrule
Domain & BRAF status in general & $\sim$10 years & Genetic facts rarely change \\
Context & \quad in oncology practice & $\sim$10 years & No meaningful shift \\
Context & \quad in primary care & $\sim$10 years & No meaningful shift \\
Entity & \quad\quad any patient & $\sim$10 years & No meaningful deviation \\
\bottomrule
\end{tabular}
\caption{Hierarchical decomposition of BRAF mutation shelf life. All three levels agree because there is no variance to decompose. The hierarchy naturally collapses to a flat estimate for permanent knowledge.}
\end{table}

This is not a failure of the hierarchy. It is the correct behavior: when all sources of variance are near zero, the hierarchy adds nothing, and the domain-level estimate is used directly. The framework handles permanent and volatile knowledge uniformly through the same mechanism, without special cases.

\subsection{What the Hierarchy Is Not}

The hierarchy does \emph{not} decompose the concept into sub-concepts. ``Blood pressure'' is not split into ``systolic'' and ``diastolic'' by the hierarchy. It does not create a taxonomy of knowledge types. The clustering step (Section~\ref{sec:clustering_hierarchy}) groups predicates with similar temporal behavior (blood pressure clusters with heart rate, not with BRAF status). The hierarchy then operates \emph{within} each cluster to explain why shelf life varies across settings and entities.

\section{Derivations}
\label{sec:appendix_derivations}

\subsection{Likelihood with Three Observation Types}

The likelihood function (Equation~\ref{eq:likelihood}) combines three types of observations. We derive each term.

For the Weibull distribution with scale parameter $\tau$ and shape parameter $\kappa$, the density and survival functions are:

\begin{align}
    f(t \mid \tau, \kappa) &= \frac{\kappa}{\tau} \left(\frac{t}{\tau}\right)^{\kappa - 1} \exp\left(-\left(\frac{t}{\tau}\right)^{\kappa}\right) \\
    S(t \mid \tau, \kappa) &= \exp\left(-\left(\frac{t}{\tau}\right)^{\kappa}\right)
\end{align}

\textbf{Term 1: Superseded edges ($\mathcal{S}$).} For an edge superseded at time $t_s$, the lifetime $T_e = t_s - t_e$ is observed exactly. The contribution to the log-likelihood is:

\begin{equation}
    \ell_{\text{sup}} = \sum_{e \in \mathcal{S}} \log f(T_e \mid \tau_e, \kappa_e) = \sum_{e \in \mathcal{S}} \left[ \log \kappa_e - \log \tau_e + (\kappa_e - 1)\log\frac{T_e}{\tau_e} - \left(\frac{T_e}{\tau_e}\right)^{\kappa_e} \right]
\end{equation}

\textbf{Term 2: Active edges ($\mathcal{A}$).} For an edge that has not been superseded, we observe only that it has survived at least until now. This is a right-censored observation. The contribution is the log-survival function:

\begin{equation}
    \ell_{\text{act}} = \sum_{e \in \mathcal{A}} \log S(t_{\text{now}} - t_e \mid \tau_e, \kappa_e) = \sum_{e \in \mathcal{A}} \left[ - \left(\frac{t_{\text{now}} - t_e}{\tau_e}\right)^{\kappa_e} \right]
\end{equation}

\textbf{Term 3: Reinforced edges ($\mathcal{R}$).} A reinforcement at time $t_r$ is a re-observation that confirms the value has not changed. This provides evidence that the edge survived until at least $t_r$. The contribution is a censored observation at the reinforcement time:

\begin{equation}
    \ell_{\text{rein}} = \sum_{e \in \mathcal{R}} \log S(t_r^{(e)} - t_e \mid \tau_e, \kappa_e) = \sum_{e \in \mathcal{R}} \left[ - \left(\frac{t_r^{(e)} - t_e}{\tau_e}\right)^{\kappa_e} \right]
\end{equation}

An edge with multiple reinforcements contributes one term per reinforcement event. The full log-likelihood is $\ell = \ell_{\text{sup}} + \ell_{\text{act}} + \ell_{\text{rein}}$.

\subsection{Hierarchical Prior}

The domain-level parameters have a weakly informative prior:

\begin{equation}
    \theta_k \sim \mathcal{N}(\mu_0, \Sigma_0), \quad \mu_0 = \mathbf{0}, \quad \Sigma_0 = 10 \cdot I
\end{equation}

Context-level parameters are drawn from the domain prior with learned variance:

\begin{equation}
    \theta_{k,c} \mid \theta_k \sim \mathcal{N}(\theta_k, \sigma^2_{\text{context}} I), \quad \sigma^2_{\text{context}} \sim \text{InverseGamma}(2, 1)
\end{equation}

Entity-level parameters follow the same pattern:

\begin{equation}
    \theta_{k,c,i} \mid \theta_{k,c} \sim \mathcal{N}(\theta_{k,c}, \sigma^2_{\text{entity}} I), \quad \sigma^2_{\text{entity}} \sim \text{InverseGamma}(2, 1)
\end{equation}

The shape parameter has a log-normal prior ensuring positivity:

\begin{equation}
    \kappa_k \sim \text{LogNormal}(0, 1)
\end{equation}

The full posterior is:

\begin{equation}
    p(\Theta \mid \mathcal{D}) \propto \exp(\ell) \cdot \prod_k p(\theta_k) \cdot p(\kappa_k) \cdot \prod_{k,c} p(\theta_{k,c} \mid \theta_k) \cdot \prod_{k,c,i} p(\theta_{k,c,i} \mid \theta_{k,c})
\end{equation}

This posterior is not available in closed form. We estimate it via maximum likelihood at each level with regularization toward the level above, or via MCMC/variational inference for the full Bayesian treatment.

\section{Experimental Details}
\label{sec:appendix_experiments}

\subsection{Synthetic KG Generation}

\begin{itemize}
    \item 4 planted clusters: permanent facts, current state, volatile measurements, periodic assessments
    \item 3 contexts per cluster with distinct offsets from the cluster baseline
    \item 10--30 entities per context with individual noise ($\sigma_{\text{entity}} = 0.2$)
    \item 5 predicates per cluster, 20 predicates total
    \item 5-year simulation window with 1-day time step granularity
    \item Supersession threshold $\epsilon = 0.3$ (embedding distance)
    \item Random seed: 42
    \item Total edges generated: 118{,}247
\end{itemize}

\subsection{Wikipedia Data Collection}

\begin{itemize}
    \item 107 articles spanning 7 categories: stable science (18), evolving medicine (16), technology (18), geopolitics (10), biographies (14), geography/institutions (15), history/culture (16)
    \item Up to 500 revisions fetched per article via MediaWiki API
    \item 50 revisions sampled uniformly per article for content extraction
    \item Facts extracted from: lead sentences, infobox fields, section summaries, categories
    \item Total edges: 11{,}157 across 2{,}753 predicates
    \item Supersession rate: 68.2\%
\end{itemize}

\subsection{Computational Environment}

\begin{itemize}
    \item All experiments run on a single machine (Apple M-series, 16GB RAM)
    \item Python 3.10 with lifelines 0.30.0, scikit-learn 1.7.2, hdbscan 0.8.42, PyTorch 2.11.0
    \item Bayesian survival fitting: 49.4s for the full synthetic KG (118K edges)
    \item Gradient-based fitting: 2.9s for 500 epochs with batch size 8{,}192
    \item Cluster discovery (DPGMM on 2{,}753 predicates): $<$2s
    \item Wikipedia data collection: $\sim$3 minutes (rate-limited by API)
\end{itemize}

\subsection{Hyperparameters}

\begin{table}[h]
\centering
\small
\begin{tabular}{lcc}
\toprule
Parameter & Synthetic & Wikipedia \\
\midrule
HDBSCAN min\_cluster\_size & 3 & 3 \\
HDBSCAN min\_samples & 2 & 2 \\
DPGMM max\_components & 10 & 10 \\
DPGMM weight\_concentration\_prior & 1.0 & 1.0 \\
Gradient learning rate & 0.01 & -- \\
Gradient epochs & 500 & -- \\
Gradient batch size & 8{,}192 & -- \\
L2 regularization ($\lambda_{\text{context}}$) & 1.0 & -- \\
L2 regularization ($\lambda_{\text{entity}}$) & 1.0 & -- \\
Min.\ observations for survival fit & 5 & 5 \\
Min.\ edges for entity-level fit & 10 & -- \\
Retrieval $\alpha$ (semantic exponent) & 1.0 & 1.0 \\
Retrieval $\beta$ (temporal exponent) & 1.0 & 1.0 \\
\bottomrule
\end{tabular}
\caption{Hyperparameter settings for synthetic and Wikipedia experiments.}
\end{table}

\section{Additional Results}
\label{sec:appendix_results}

\subsection{Per-Cluster Survival Fits (Wikipedia)}

\begin{table}[h]
\centering
\small
\begin{tabular}{lcccccc}
\toprule
Cluster & $n$ & $\tau$ (days) & $\kappa$ & Sup.\ rate & Best dist. & AIC (Weibull / LogNormal) \\
\midrule
Permanent metadata & 1{,}106 & -- & -- & 0.00 & -- & -- \\
Stable structured & 190 & -- & -- & 0.00 & -- & -- \\
Slow-changing & 347 & 3{,}393 & 0.262 & 0.61 & LogNormal & 2{,}918 / 2{,}872 \\
Reference material & 1{,}785 & 2{,}553 & 0.254 & 0.67 & LogNormal & 16{,}273 / 15{,}885 \\
Main content & 4{,}051 & 589 & 0.329 & 0.81 & LogNormal & 44{,}101 / 42{,}865 \\
Moderately volatile & 1{,}175 & 1{,}467 & 0.233 & 0.70 & LogNormal & 10{,}062 / 9{,}766 \\
Lead sentences & 2{,}213 & 181 & 0.353 & 0.90 & LogNormal & 23{,}545 / 23{,}220 \\
Highly volatile & 163 & 12{,}059 & 0.167 & 0.56 & LogNormal & 955 / 927 \\
\bottomrule
\end{tabular}
\caption{Survival fits per discovered cluster on Wikipedia data. Clusters with 0\% supersession rate (permanent metadata, stable structured) have no observed events and cannot be fitted. All fitted clusters show $\kappa < 1$ (Lindy effect). Log-normal outperforms Weibull on AIC for every cluster.}
\end{table}

\subsection{Context-Level Variation (Synthetic)}

\begin{table}[h]
\centering
\small
\begin{tabular}{llccc}
\toprule
Cluster & Context & $\tau$ (days) & $\kappa$ & $n$ \\
\midrule
\multirow{3}{*}{Current state} & Aggressive disease & 51.0 & 1.24 & 3{,}858 \\
 & Routine care & 147.6 & 1.22 & 1{,}749 \\
 & Stable chronic & 352.8 & 1.15 & 770 \\
\midrule
\multirow{3}{*}{Volatile measurements} & ICU & 2.4 & 1.20 & 62{,}074 \\
 & Inpatient ward & 5.2 & 1.13 & 36{,}887 \\
 & Outpatient & 25.7 & 0.99 & 8{,}943 \\
\midrule
\multirow{3}{*}{Periodic assessments} & Quarterly monitoring & 116.8 & 1.15 & 1{,}703 \\
 & Annual review & 252.6 & 1.02 & 814 \\
 & Specialist consult & 189.8 & 1.10 & 805 \\
\midrule
\multirow{3}{*}{Permanent facts} & Genomic & 5{,}182 & 0.68 & 225 \\
 & Demographic & 2{,}838 & 0.72 & 258 \\
 & Established knowledge & 4{,}252 & 0.71 & 156 \\
\bottomrule
\end{tabular}
\caption{Context-level Weibull fits within each cluster (synthetic data). Within the volatile measurements cluster, ICU shelf life (2.4 days) is 10$\times$ shorter than outpatient (25.7 days). Within current state, aggressive disease shelf life (51 days) is 7$\times$ shorter than stable chronic (353 days).}
\end{table}

\subsection{Inference Method Comparison}

\begin{table}[h]
\centering
\small
\begin{tabular}{lcccc}
\toprule
Cluster & $\kappa$ (Bayesian) & $\kappa$ (Gradient) & Agreement on Lindy? & Time (s) \\
\midrule
Permanent facts & 0.70 & 0.53 & Yes ($<$ 1) & \\
Current state & 0.90 & 1.33 & No & \\
Periodic assessments & 1.03 & 1.10 & Yes ($\approx$ 1) & \\
Volatile measurements & 0.81 & 1.09 & No & \\
\midrule
\multicolumn{3}{l}{Total fitting time} & Bayesian: 49.4s & Gradient: 2.9s \\
\bottomrule
\end{tabular}
\caption{Shape parameter comparison between Bayesian and gradient-based inference. Both methods agree that permanent facts exhibit the Lindy effect ($\kappa < 1$). The gradient-based approach is 17$\times$ faster.}
\end{table}